\documentclass[useAMS]{mn2e}

\usepackage{amssymb,amsmath,psfig,times}
\usepackage{graphicx}
\def\gsim{ \lower .75ex \hbox{$\sim$} \llap{\raise .27ex \hbox{$>$}} }
\def\lsim{ \lower .75ex\hbox{$\sim$} \llap{\raise .27ex \hbox{$<$}} }

\def\sc{Schwarzschild}
\def\beq{\begin{equation}}
\def\eeq{\end{equation}}

\def\sc{Schwarzschild}

\def\sw{{\it Swift}}
\def\fe{{\it Fermi}}
\def\ba{BATSE}

\def\ep{$E_{\rm peak}$}

\def\epo{$E^{\rm obs}_{\rm peak}$}

\def\liso{$L_{\rm iso}$}
\def\eiso{$E_{\rm iso}$}
\def\ama{$E_{\rm peak}-E_{\rm iso}$}
\def\yone{$E_{\rm peak}-L_{\rm iso}$}

\def\flim{$F_{\rm lim}$}

\title[\yone\ correlation]
{The impact of selection biases on the \yone\ correlation of Gamma Ray Bursts}
\author[G. Ghirlanda et al.]
{G. Ghirlanda$^{1}$\thanks{E--mail:giancarlo.ghirlanda@brera.inaf.it}, 
G. Ghisellini$^{1}$, L. Nava$^{2}$, R. Salvaterra$^{3}$, G. Tagliaferri$^{1}$, S. Campana$^{1}$,\newauthor
S. Covino$^{1}$, P. D'Avanzo$^{1}$, D. Fugazza$^{1}$, A. Melandri$^{1}$, S. D. Vergani$^{1}$\\
$^{1}$INAF -- Osservatorio Astronomico di Brera, via E. Bianchi 46, I-23807 Merate, Italy\\
$^{2}$APC Universit\'e Paris Diderot, 10 rue Alice Domon et Leonie Duquet, F-75205 Paris Cedex 13, France\\
$^{3}$INAF - IASF Milano, via E. Bassini 15, I-20133 Milano, Italy \\
}
\begin{document}

\date{}


\maketitle

\label{firstpage}

\begin{abstract}
We study the possible effects of selection biases on the \yone\ correlation caused by the unavoidable 
presence of  flux--limits in the existing samples of Gamma Ray Bursts (GRBs). 
We consider a well defined complete sample of \sw\ GRBs and perform Monte Carlo 
simulations of the GRB population under different assumptions for their luminosity functions. 
If we assume that there is no correlation between the peak energy \ep\ and the isotropic luminosity \liso, we 
are unable to reproduce it as due to the flux limit threshold of the \sw\ complete sample.  
We can reject the null hypothesis that there is no intrinsic correlation 
between \ep\ and \liso\ at more than 2.7$\sigma$ level of confidence. This result is robust against 
the assumptions of our simulations and it is confirmed if we consider, instead of \sw,  
the trigger threshold of the \ba\ instrument. Therefore, there must be a physical relation between these two quantities.
Our simulations seem to exclude, at a lower confidence level of 1.6$\sigma$, the possibility that the observed \ep--\liso\
correlation among different bursts is caused by a boundary, i.e.  such that for any given \ep, we see only the largest
\liso, which has a flux above the threshold of the current instruments.
\end{abstract}

\begin{keywords}
Gamma-ray: bursts  --- Radiation mechanisms: non thermal
\end{keywords}

\section{Introduction}

The isotropic luminosity \liso\  and the isotropic energy \eiso\ of Gamma Ray 
Bursts (GRBs) are  strongly correlated with their peak spectral energy \ep\ 
(i.e. the peak of the $\nu F_{\nu}$ spectrum). 
The spectral--energy correlations of GRBs gathered the interest of the 
scientific community for their possible implications on the physics of GRBs 
(Yamazaki, Ioka \& Nakamura 2004; Eichler \& Levinson 2005; 
Lamb, Donaghy \& Graziani 2005; Levinson \& Eichler 2005;
Rees \& Meszaros 2005; Toma, Yamazaki \& Nakamura 2005;
Barbiellini et al. 2006; Ryde et al. 2006; Thompson 2006; 
Giannios \& Spruit 2007; Thompson, Meszaros \& Rees 2007; 
Guida, Bernardini \& Bianco 2008; Panaitescu 2009),
but also raised an intense debate about the strong selection biases 
they might suffer of 
(Band \& Preece 2005; Nakar \& Piran 2005; Butler et al. 2007; 
Butler, Kocevski \& Bloom 2009; Shahmoradi \& Nemiroff 2011; 
Kocevski 2012; but see Ghirlanda, Ghisellini \& Firmani 2005; 
Bosnjak et al. 2008; Ghirlanda et al. 2008; Nava et al. 2008; 
Amati, Frontera \& Guidorzi 2009; Krimm et al. 2009). 

Ghirlanda et al. (2012) studied the comoving frame properties of a 
small sample of GRBs for which the bulk Lorentz factor $\Gamma$ could be estimated. 
They found that $\Gamma$ is correlated with $L^2_{\rm iso}$ and with \ep. 
These newly found correlations offer an interpretation of the \yone\ correlation 
as a sequence of $\Gamma$ factors, the most luminous GRBs have the largest $\Gamma$ and \ep. 

Since these correlations involve the rest frame prompt $\gamma$--ray emission properties 
(\ep, \eiso\ and \liso) of GRBs with measured redshift $z$, it has been argued that 
they are strongly biased by 
(i) the instrumental detector--trigger threshold and/or by 
(ii) the redshift dependence of the involved observables (e.g. Butler et al. 2007). 

Different studies  (Ghirlanda et al. 2008; Nava et al. 2008) quantified the possible 
instrumental selection biases finding that, even if present, 
they cannot be responsible for the existence of the spectral--energy correlations. 
While the \ama\ and \yone\ correlations are defined considering the time--integrated 
spectra of GRBs, it has been shown (Firmani et al. 2009; Ghirlanda et al. 2010; 2011a; 2011b) 
that similar correlations hold within individual bursts. 
This is a strong argument in favour of the physical origin of the correlations since
within a single burst the instrumental selection effects 
(e.g. the flux--limited threshold of any burst detector), or possible effects due to 
the dependence of \ep, \eiso\ and \liso\ on the redshift do not play a role. 

Discovered with a dozen of GRBs with known redshifts (Yonetoku et al. 2004; Amati et al. 2002), 
the \yone\ and the \ama\  correlations have been updated in the last decade by an increasing 
number of bursts with measured $z$ (see e.g. Nava et al. 2011 for a recent update). 
Nevertheless, these samples are not complete in the sense that there is no well defined
flux limited sample with the redshifts measured for all GRBs.
This raised the suspect that the incompleteness in the 
redshift knowledge might bias such correlations. 

Nava et al. 2012 (N12, hereafter) studied the \ama\ and the \yone\ correlations with a flux 
limited sample of GRBs detected by \sw\ which has also a high level of redshift determination. 
This complete \sw\ sample, presented in Salvaterra et al. 2012 (S12, hereafter), was 
selected starting from the 50\% redshift complete sample of 
Jackobsson\footnote{http://www.rauvis.hi.is/$\sim$pja/GRBsample.html} and considering the 
58 bright \sw\ GRBs with a 1 s peak flux (integrated in the 15--150 keV energy band) 
$F\ge F_{\rm lim}=2.6$ ph cm$^{-2}$ s$^{-1}$. 
The redshift recovery rate of a such selected sample turns out to be 90\%. 
Therefore, this flux--limited sample has also a high level of completeness in redshift. 
S12 study the luminosity function (LF) of GRBs throughout the \sw\ complete sample finding 
that evolution in luminosity or in density is required in order to account for the observations. 
Among other results, N12 find  that the  correlations defined with the complete \sw\ sample 
are statistically robust: the rank correlation coefficient is $\rho=0.76(0.70)$ and chance 
probability $P=7\times10^{-10}(1\times10^{-6})$ for the \ama\ (\yone) correlation. 
Moreover, the correlation properties (slope and normalization) of the complete \sw\ sample 
are consistent with those defined with the incomplete larger sample of 136 bursts with known 
$z$ and spectral parameters (see N12).

Complete samples are at the base of any population studies. 
In this paper we study the impact of instrumental selection effects on the \yone\ correlation 
using as a reference the complete \sw\ sample of GRBs described in S12 and analyzed 
(in terms of its \yone\ correlation) in N12. 
In particular, since no clear consensus has been  reached yet on the physical origin of 
these correlations  we aim to answer to this specific question: {\it might the \yone\ correlation be 
produced by the threshold of a flux--limited sample of bursts?}

To this aim we use a population synthesis code that generates a large sample of GRBs 
(following some prescriptions for its luminosity and redshift distribution -- described in \S2) 
and {\it assuming that there is no correlation between \ep\ and \liso}. 
Throughout the paper we make use of the following nomenclature: 
\begin{itemize}

\item ``complete \sw\ sample": this is the sample defined in S12 of 58 GRBs detected by 
\sw--BAT with $F\ge$ \flim, of which 90\% have measured redshifts.  Forty--six out of 
58 bursts (79\%) have well constrained peak energy and \liso\ and their \yone\ 
correlation (which is the term of comparison of our simulations) has been published in N12. 
The other 12 GRBs are consistent with the \yone\ correlation as discussed in N12. 

\item ``complete sample of simulated bursts": this is the sub--sample of simulated GRBs with 
$F\ge$\flim, i.e. representative of the ``complete \sw\ sample''. 

\end{itemize}
By applying the \sw\ flux limit  \flim\ (i.e. the same used to define the complete \sw\ sample in S12), 
we test if a strong \yone\ correlation is obtained in the complete sample of simulated bursts 
(\S3) as only due to the cut in the flux.  
If we cannot find a statistically significant \yone\ correlation in the simulated complete 
sample we then reject the null hypothesis that the real one  is due to the flux--limit of 
the sample selection. 
In the latter case, we are left with two possibilities, that we also test in this work (\S5), 
for the nature of the \yone\ correlation: either it is an intrinsic correlation to GRBs 
(as also supported by the existence of a similar correlation within individual 
bursts -- Ghirlanda et al. 2010, 2011, 2011a) with a symmetric scatter of data points around it, or it is 
 caused by a boundary (as proposed by Nakar \& Piran 2005), i.e. there is a considerable fraction of bursts with 
intermediate/large \ep\ and low luminosities (\S5). 
We also verify with our code (\S4) the recent claims on selection effects (Kocevski 2012) 
induced by the \ba\ trigger threshold.

The main advantage of our population synthesis code is that it relies on a small number 
of assumptions (described in \S2). 
We focus on testing the \yone\ correlation because for any simulated GRB with given $z$, \liso\ and 
\ep, its 15--150 keV peak flux is easily compared with the \flim\ of the \sw\ complete sample. 
The adoption of the complete \sw\ sample is also relevant, because the \yone\ correlation 
defined with this sample is independent from biases induced by the measurement of the redshifts (N12). 
Throughout the paper we assume a standard flat universe with $h=\Omega_{\Lambda}=0.7$. 

\section{Simulation setup} 

We need to simulate a population of bursts with $z$ and \liso\ assuming a redshift density 
distribution and a luminosity function, respectively. To every simulated GRB we assign also 
a peak energy \ep\ and assume a typical spectral shape in order to compute its 15--150 keV 
flux and compare it with \flim. 
The main steps of our population synthesis code are:
\begin{enumerate}
\renewcommand{\theenumi}{(\arabic{enumi})} 

\item We simulate a population of GRBs distributed in redshift $z$ according to the 
GRB formation rate (GRBFR)  $\psi(z)$ derived by Li 2008 (which extended to higher 
redshifts the results of Hopkins \& Beacom 2008):
\begin{equation}
\psi(z)=\frac{0.0157+0.118z}{1+(z/3.23)^{4.66}}
\label{z}
\end{equation}
where $\psi(z)$ is in units of ${\rm M}_{\sun}$ ${\rm yr}^{-1}$  ${\rm Mpc}^{-3}$. 
This is the same $\psi(z)$ adopted by S12 to derive the LF from the complete \sw\ sample. 
We simulate bursts with $z\le10$.

\item We assign peak luminosities to the simulated GRBs adopting the same luminosity function 
$\phi(L_{\rm iso})$ used  by S12 for studying the complete \sw\ sample:
\begin{eqnarray}
\phi(L_{\rm iso})&\propto&\left(L_{\rm iso}\over L_{\rm cut}\right)^{a}; \quad
                            L_{\rm iso}\le L_{\rm cut} \nonumber\\
	             &\propto& \left(L_{\rm iso}\over L_{\rm cut}\right)^{b}; \quad
	                        L_{\rm iso}> L_{\rm cut}
\label{lf}
\end{eqnarray}
In S12 both the cases of a luminosity function evolving with redshift as 
$(1+z)^{\delta_{\rm L}}$ or a density evolution 
of the GRB population as $(1+z)^{\delta_{\rm D}}$ are considered.  
Therefore, we distinguish these two cases.
Note that S12 assumed the \yone\ correlation in  constraining the LF free parameters 
$(a, b, L_{\rm cut}, \delta_{\rm L}, \delta_{\rm D})$. 
Since this is in contrast with the null hypothesis we aim to test (i.e. that there is no 
\yone\ correlation), we cannot use the parameter values of the LF reported in S12. 
For consistency with our null hypothesis, we have recomputed the LF of the complete \sw\ sample with 
the same analytic method adopted in S12 but without assuming the \yone\ correlation. 
The LF parameters $(a,b,L_{\rm cut},\delta_{\rm L}, \delta_{\rm D})$ adopted in our 
simulation (see Tab. \ref{tab1}) are consistent, within the errors, with those reported in S12. 

\item We assign to each simulated burst a peak energy \ep. 
We consider that \ep\ is uncorrelated with \liso. 
We adopt a log--normal $\xi(E_{\rm peak})$ distribution centered at 337 keV 
(i.e. the average value for long GRBs with known $z$) with a dispersion of $\sigma=0.6$ dex. 
Assuming such a large dispersion of $\xi(E_{\rm peak})$ ensures  that we simulate soft GRBs 
(i.e. X--ray flashes) with \ep\ of few tens of keV and hard bursts with \ep\ of tens of MeV. 

\item We derive the  peak
flux of each simulated burst in the 15--150 keV energy band of \sw--BAT: 
\begin{equation}
F=\int_{15\, \rm keV}^{150\, \rm keV} N(E;\alpha,\beta,E^{\rm obs}_{\rm peak},z)dE
\end{equation}
where $N(E;\alpha,\beta,E^{\rm obs}_{\rm peak},z)$ is the observer frame photon spectrum re-normalized 
through \liso. We assume for the GRB spectra the Band function (Band et al. 1993) with fixed low and high 
energy spectral index, $\alpha=-1.0$ and $\beta=-2.25$ (as done in S12).

%
\end{enumerate} 

From the simulated sample of GRBs we extract a sub--sample of bursts with (a) $F\ge$\flim$=2.6$ ph cm$^{-2}$s$^{-1}$ and 
(b) \epo\ within 15 keV and 2 MeV. The GRBs used for the definition of the spectral--energy correlations 
have their \epo\ measured. It is not mandatory that \epo\ is measured through \sw\ data, since most of the
\sw\ bursts in the complete sample have been observed by the Konus Wind satellite
or other instruments (such as the GBM onboard {\it Fermi}). Still we require that \epo\ is within the range of 
energy covered by the present instruments. The sub--sample selected as above is 
the ``complete sample of simulated GRBs". 

We then study the correlation between \ep\ and \liso\ of the complete sample of simulated bursts 
and compare it with the  observed correlation defined by the complete \sw\ sample.  We will also perform the same simulation 
by modifying assumption (3). In particular we will assume that \ep\ and \liso\ are correlated as found in real GRBs   of the complete \sw\ sample (\S5.1) or that 
the \yone\ correlation is  caused by a boundary in the corresponding plane (\S5.2).

\begin{figure}
\hskip -0.6truecm
\psfig{file=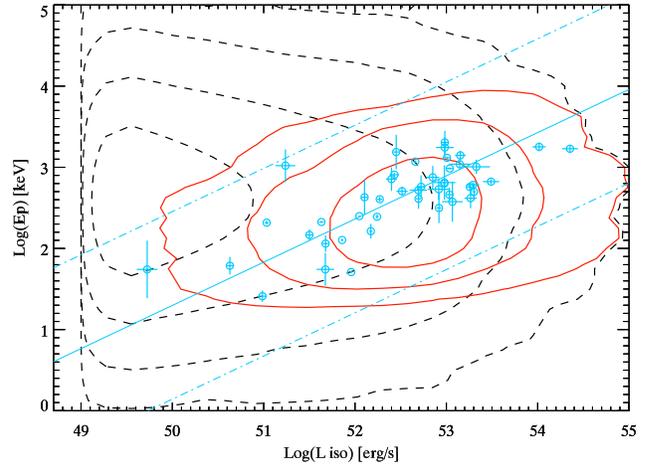,width=9.5cm}
\caption{
Simulation of GRBs with density evolution (see Tab. \ref{tab1} for the parameters). 
The simulated population of GRBs is shown by the dashed contour (1, 2, 3 and 4 $\sigma$ levels). 
The complete sample of simulated GRBs, i.e. those with a flux in the 15--150 keV energy range 
larger than the flux limit \flim=2.6 ph cm$^{-2}$ s$^{-1}$ of the \sw\ complete sample, is shown by the 
(red) solid contours (1, 2 and 3 $\sigma$ levels). 
The open (blue) circles are the 46 GRBs of the complete \sw\ sample analyzed in N12: 
the solid line is the best fit \yone\ correlation to these bursts and the dot--dashed lines 
represent the 3$\sigma$ scatter of the data points around the best fit line.  
}
\label{fg1}
\end{figure}
\begin{figure}
\hskip -0.6truecm
\psfig{file=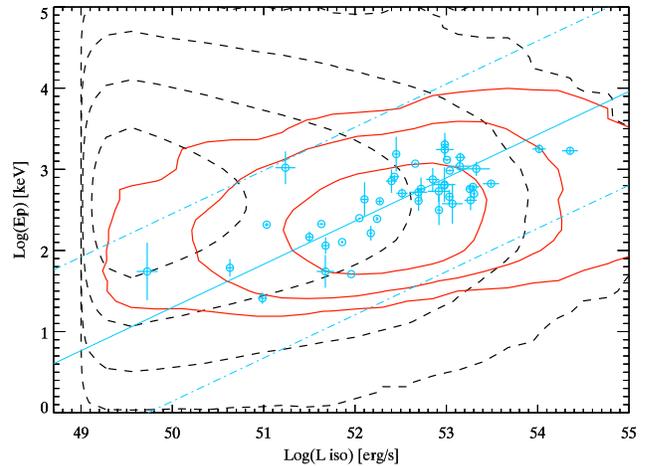,width=9.5cm}
\caption{Simulation of GRBs with luminosity evolution (see Tab. \ref{tab1} for the parameters). 
Same symbols as in Fig. \ref{fg1}. }
\label{fg2}
\end{figure}

\begin{table*}
\begin{center}
\begin{tabular}{llllllllllll}
\hline
\hline
$\Phi(L)$                & $a$          & $L_{\rm cut}$ erg s$^{-1}$         & $b$	        & $\delta$    &\vline & $\mathcal{P}_{\rho}$        & $\mathcal{P}$     	             &  \%Out. $\uparrow$ & \%Out. $\downarrow$             		\\
\hline
Density 	                 & -1.37        & 3.8$\times10^{52}$ & -2.37          & 1.22         &\vline & 7.3\%                                  & 0.7\%                  & 0.7\%	 &  2.0\% \\
Luminosity             & -1.4          & $10^{51}$                & -2.13          & 2.67         &\vline & 8.3\%                                 &  0.6\%                  & 1.0\% & 2.2\% \\
Assume \yone\      & -1.4          & $10^{51}$                & -2.13          & 2.67         &\vline & 100\%                                &  66\%		            & 0.07\% &  0.2\% \\
\yone\ boundary    &  -1.4         & $10^{51}$                & -2.13          & 2.67         &\vline & 87\%                                  &  12\%                   & 1.4\% & 0.1\%  \\
\hline
K12 (BATSE)         & -1.22	    & $10^{53}$                & -3.89         &                  &\vline & 0.5\%					&  0.0\%	                   & 2.6\% & 0.7\% \\
K12 (Swift)		    & -1.22	    & $10^{53}$                & -3.89         &                  &\vline & 0.7\%					&  0.0\%				&0.4\% & 1.3\% \\			  
\hline
\end{tabular}
\caption{
$\mathcal{P}_{\rho}$ is the percentage of simulations giving a significant correlation (i.e. with chance probability $\le 10^{-3}$). 
$\mathcal{P}$ is the percentage of simulations giving a significant correlation with slope $m$ and normalization $q$ consistent, within their 1$\sigma$ errors,  
with the correlation of the real \sw\ complete sample and with a scatter $\sigma_{c}\le 0.29$ (i.e. that of the observed correlation of the \sw\ complete sample). 
\% Out $\downarrow$ 
($\uparrow$) give the percentage of GRBs  in the complete sample of simulated bursts which 
are outliers at more  than 3$\sigma$ of the \yone\ correlation defined by the complete \sw\ sample. 
The arrows correspond to the outliers below ($\downarrow$) and above ($\uparrow$) the boundary 
of the 3$\sigma$ scatter (dot--dashed blue lines in all the figures). 
}
\label{tab1}
\end{center}
\end{table*}

\section{Results}

By studying the complete \sw\ sample of GRBs, N12 found that 
the \yone\ correlation defined by the 46 GRBs in this sample with known \ep\ and \liso\ has the following 
properties:
\begin{itemize} 
\item its rank correlation coefficient is $\rho=0.7$ and its associated chance probability $P_{\rho}=10^{-6}$, i.e. 
it is significant at more than 3$\sigma$;
\item the fit of the correlation with a powerlaw $\log(E_{\rm peak})=m\log(L_{\rm iso})+q$ 
gives $m=0.53\pm0.06$ (1$\sigma$) and $q=-25.3\pm3.2$ (1$\sigma$) for the correlation slope and normalization, respectively;
\item the scatter of the 46 GRBs computed perpendicular around this best fit line has a dispersion of
$\sigma_{\rm c}\simeq0.29$ dex;
\item the complete \sw\ sample has no outlier  at more than 3$\sigma$, except for one burst at the limit of the 
3$\sigma$ dispersion of the correlation.
\end{itemize} 

The goal of our simulations, for a given set of input assumptions (i.e. $\psi(z)$, $\phi(L_{\rm iso})$, $\xi(E_{\rm peak}$), 
is (1) to produce a significant correlation in the \yone\ plane and (2) to obtain a correlation which is consistent with that of the 
real GRB sample of comparison, i.e. the \sw\ complete sample. The simulation described in \S2 is repeated  300 times 
with the same initial assumptions. For each  repeated simulation the complete sample of simulated GRBs is analyzed deriving the parameters of its \yone\ correlation. 
At the end of a cycle,  i.e. after 300 repeated simulations, we derive:
\begin{enumerate}

\item the percentage $\mathcal{P}_{\rho}$ (Col. 6 in Tab. \ref{tab1}) of simulations showing a \yone\ correlation of the complete sample of 
simulated GRBs significant at least at the 3$\sigma$ level. We consider a correlation significant if the chance probability of its 
rank correlation coefficient is $\le10^{-3}$. 

\item the percentage $\mathcal{P}$ (Col. 7 in Tab. \ref{tab1}) of simulations significant at more than $3\sigma$ with a slope $m$ and normalization $q$ consistent (within their 1$\sigma$ errors) 
with the slope and normalization of the correlation defined by the \sw\ complete sample and with a scatter $\sigma_{\rm c}\le 0.29$, i.e. that of the observed correlation;

\item the average percentage of simulated bursts which are outliers at more than 3$\sigma$ of the correlation defined by
the \sw\ complete sample (Col. 8 and 9 in Tab. \ref{tab1}). 
\end{enumerate}

For a certain set of assumptions ($\psi(z)$ and $\phi(L_{\rm iso})$) our term of comparison is the correlation observed
in the \sw\ complete sample (N12). Therefore, the best match between the ``simulated world''
and the ``reality'' is when  $\mathcal{P}_{\rho}$ and $\mathcal{P}$ are large and the percentage of outliers  at more than 3$\sigma$ is 
consistent with 0.3\% 

In addition, we also check that the complete sample of simulated GRBs has a redshift distribution 
and a flux distribution consistent with those of the real complete  \sw\  sample.  
Since we adopt the LF derived from the complete sample of S12, we should find a complete 
sample of simulated bursts with $\psi(z)$ and $\phi(L_{\rm iso})$ consistent with those of the real sample, and indeed this is the case. 

Tab. \ref{tab1} lists the input assumptions (LF parameters) and  the results of our simulations which are also 
shown in Fig. \ref{fg1}, \ref{fg2}, \ref{fg3}, \ref{fg2a}, \ref{fg4}: the contours are 
obtained by smoothing the distribution of data in the \yone\ plane obtained by stacking all 
the  300 repeated simulations. The red contours, for instance, are obtained by smoothing the distribution of 13800 data points, i.e. 
46 simulated GRBs in 300 repeated simulations.

\section{Null hypothesis: no intrinsic \yone\ correlation}

Under the null hypothesis that there is no intrinsic correlation between \ep\ and \liso\ we find that $\sim$92.7\% (91.7\%) of the 300 repeated simulations do not 
produce a significant (at least at the 3$\sigma$ level of confidence) \yone\ correlation (Tab. \ref{tab1}) in the case of density (luminosity) evolution.   
Moreover, if we also require that the correlation obtained with the simulated bursts is consistent (in terms of its slope, normalization and 
scatter) with that observed in the \sw\ complete sample, the percentage of simulations satisfying all our constraints drops to $\mathcal{P}=0.7$\% ($\mathcal{P}=0.3$\% in the case of luminosity evolution). 
In other words, we can never reproduce the observed correlation through our simulations if we assume that \ep\ and \liso\ are uncorrelated. 
Therefore, we can reject the null hypothesis (i.e. that there is no correlation between \ep\ and \liso) at the 2.7$\sigma$ and 3$\sigma$ confidence level in the case of density and luminosity evolution of the GRB population, respectively. These results are shown in Figs. \ref{fg1}, \ref{fg2} where the solid contours represent the sub--sample of simulated bursts 
extracted with the same flux limit of the \sw\ complete sample (open blue 
circles in these figures) and the dashed contours represent the total population of simulated bursts. From these plots, it appears that 
the simulated bursts cannot reproduce the distribution of the real data points (blue open circles) 
in the \yone\ plane.  
 
The solid (red) contours of Fig. \ref{fg1} show a very weak correlation (in the upper--left part of the plane). Indeed, the flux--limit induces a weak 
correlation in the \yone\ plane by excluding part of the observables' space (i.e. the upper--left part of the \yone\ plane). However, a very weak correlation 
as that shown by the (red) solid lines in Fig. \ref{fg1}, has a very high significance if it is computed for all the 13800 data points resulting from the 300 repeated simulations. This is because the correlation significance would be dominated by the extremely large data sample. Instead, our goal is to compare samples of comparable sizes: the 46 GRBs of the swift complete sample versus the 46 GRBs of the simulated complete sample (the latter are produced in each of the 300 repeated simulations). For this reason, in analyzing our results, we have computed, for instance, the percentage of repeated simulations that generate a sample 
of 46 simulated GRBs which have a significant correlation in the \yone\ plane. 
As we discuss in \S3, although an apparent very weak correlation is produced by the flux--limit in Fig. \ref{fg1} in the uppper--left part of the plane, the 
overall distribution of the data points violates most of our constraints. 

Under the null hypothesis of no correlation between \ep\ and \liso, we have computed the average percentage 
of GRBs which are outliers at more than 3$\sigma$ below (\%Out $\downarrow$ in Tab. \ref{tab1}) the  observed \yone\ correlation defined by the \sw\ complete sample. 
According to our simulations, \sw\ should have detected 
a percentage between 2.0--2.2\% (see Tab. \ref{tab1}) of GRBs with intermediate \ep\ and 
large \liso\ (e.g. $10^{52-54}$ erg s$^{-1}$). 
These bursts (1 event on average) should be below the 3$\sigma$ limit of the \yone\ 
correlation (dot--dashed line in Fig. \ref{fg1}, \ref{fg2}) defined by the complete \sw\ sample. 
At present, there are no outliers in the bottom/right triangle of the \yone\ plane, 
 either in the complete \sw\ sample or more generally in the whole population of GRBs with measured 
$z$ and known \ep\ and \liso\ (e.g. see N12). 

A similar argument applies to the simulated GRBs of the complete sample that have high \ep\ and low \liso. 
The 0.7--1.0\% of simulated bursts with $F\ge$\flim\ lie above 
the 3$\sigma$ upper boundary of the \yone\ correlation of the real burst sample. 
While there are no outliers in the complete \sw\ sample in this part of the plane, it has been 
discussed in the literature if a fraction of bursts with intermediate/large \ep\ and low \liso\ 
could already be present in the populations of GRBs detected by different satellites 
(e.g Nakar \& Piran 2005; Band \& Preece 2006). 
On the other hand, Nava et al. (2011) have shown that there are no outliers of the \yone\ 
correlation at more than 3$\sigma$ in the \ba\ and \fe\ population of GRBs. 

Note also that a considerable fraction ($\sim$20\%) of the simulated GRBs with $F\ge$\flim\ have 
a peak energy \epo, in the observer rest frame,  larger than the upper \sw\ energy threshold of 350 keV
but still lower than 2 MeV, which roughly corresponds to the limit of current instruments with the largest 
energy range that can measure \epo\ (e.g. \fe\ and {\it Konus}). 

\begin{figure}
\hskip -0.6truecm
\psfig{file=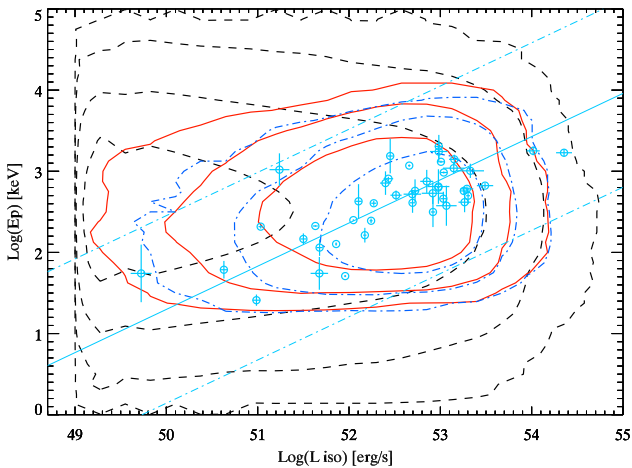,width=9.5cm}
\caption{
Results of the simulation of a GRB population with $\phi(L_{\rm iso})$ and $\xi(z)$ 
derived by Butler et al. 2011 (used by K12) and adopting the \ba\ trigger threshold 
(solid contours) or the \sw\ \flim (dot--dashed contours). 
The null hypothesis is that there is no intrinsic \yone\ correlation. 
Symbols as Fig. \ref{fg1}.  
}
\label{fg3}
\end{figure}

Kocevsky 2012 (K12, hereafter) performed a population study finding that the \ama\ correlation is induced by a combination of the 
redshift/luminosity function of GRBs with the detection limit of a given instrument (\ba\ in his study). 
K12 attributes the existence of the \ama\ correlation to the 
combination of the Malmquist bias (i.e. the detection of the most luminous 
GRBs at high $z$ preferentially), the flux limit of the detector, and its limited band--pass. 
According to this interpretation the most luminous GRBs at high $z$ with intermediate \ep\ 
goes undetected because their \epo=\ep/$(1+z)$ falls below the low energy threshold of the 
detector (i.e. 15 keV in the case of \sw--BAT). This  argument is intuitively plausible:  
a detector sensitive in a given energy range should hardly detect bursts with \epo\ outside this range. 
However, what matters for triggering a burst is its peak flux (integrated in the 
detector energy range) which depends on the spectral shape of the burst 
(e.g. \epo, $\alpha$, $\beta$) and the normalization of the spectrum. 
If this flux is larger than the detector flux limit, the burst will be detected no matter its \epo.  
Our simulations in the case of luminosity evolution, for instance, predict that $\sim$13\% 
of the simulated GRBs with $F\ge$\flim\ have \epo$<$15 keV (i.e. the lower threshold of the 
BAT energy band--pass). 
In other words, these bursts are bright enough to be detected by \sw, despite the fact 
that their \epo\ is outside the BAT energy range. 
Therefore, the absence of GRBs in the lower/right plane of the \yone\ correlation cannot 
be  entirely attributed to this effect.

Our population synthesis code is simpler than that of K12 because we  test the \yone\ 
correlation through the complete \sw\ sample. 
In our case, by simulating $z$, \liso\ and \ep\  we can immediately compute the 
peak flux of a simulated GRB  and compare it with the flux limit. The simulation of K12 concerns the \ama\ correlation,  
which involves the time integrated spectral properties of GRBs, and requires several assumptions (e.g. on the profile structure 
and time evolution of the spectrum during a burst) in order to simulate 
a GRB light curve and verify if it can be detected by \ba\ (in turn, this also requires  K12
to model  a typical background and the \ba\ detector response matrix).  
The advantage of our simulation is that it uses the \flim\ of the complete sample of 
S12, which is a simple sharp cut in the 15--150 keV peak flux of the simulated GRB population. 

K12 adopts a slightly different LF and performs the simulation for \ba. 
We assumed his luminosity function and redshift density distribution 
$\phi(L_{\rm iso})$ and $\xi(z)$ (originally derived from Butler et al. 2011). 
We have implemented in our code the \ba\ detection algorithm as described in Band (2003). 
The results of this simulation are shown in Fig. \ref{fg3} and reported in Tab.  \ref{tab1} (labelled K12). 
Under the null hypothesis of no correlation between \ep\ and \liso, the majority (99.5\%) 
of the repeated simulations show no significant \yone\ correlation in the sample of 
bursts that should be detected by \ba. The percentage of outliers that we find is 0.7\% below the correlation and 2.3\% 
above the correlation (see Fig. \ref{fg3}  and Tab. \ref{tab1}). 
We compare the results of the simulation assuming the \ba\ trigger threshold with 
the complete \sw\ sample instead of using the larger (but highly incomplete) 
sample of all the bursts with measured $z$ and known spectral parameters. 
This is justified by the findings of N12 that show that
the \yone\ correlation defined by the complete \sw\ sample (46 events with \ep\ and \liso\ known) 
is  quite similar to the correlation defined with the larger sample of 136 GRBs 
with measured $z$ and known spectral parameters (\ep, \liso).
We have also tested the LF adopted by K12 but with the flux limit of \sw. 
We show in Fig. \ref{fg3} (see also Tab. \ref{tab1}) the contours corresponding 
to the complete sample of simulated GRBs (dot--dashed contours). 
Also in this case we are not able to obtain a significant (at more than 3$\sigma$) 
\yone\ correlation in 96\% of the repeated simulations under the null hypothesis 
that there is no \yone\ correlation.

\section{\yone\ correlation or boundary?}

We have shown in \S4 that if there is no correlation between \ep\ and \liso\  the apparent correlation between these two observables 
{\it cannot be produced} by the flux limit cut of the complete \sw\ sample or of the \ba\ instrument. We still have 
two possibilities: 
(i) to assume the observed correlation and use it in simulating the GRB population; 
(ii) assume that the \yone\ correlation is  produced by a boundary in the corresponding plane. 


 
For these reasons, it is  worthwhile to test the possibility
that there is a population of GRBs with a uniform distribution 
of \ep\ above the \yone\ correlation. 
In this case their absence in the observed \yone\ plane could be induced by the trigger threshold and/or by a bias 
 related to the detection of the bursts with the smallest jet opening angles. 
Here we study the former possibility (i.e. absence due
to the detector threshold).

\begin{figure}
\hskip -0.6truecm
\psfig{file=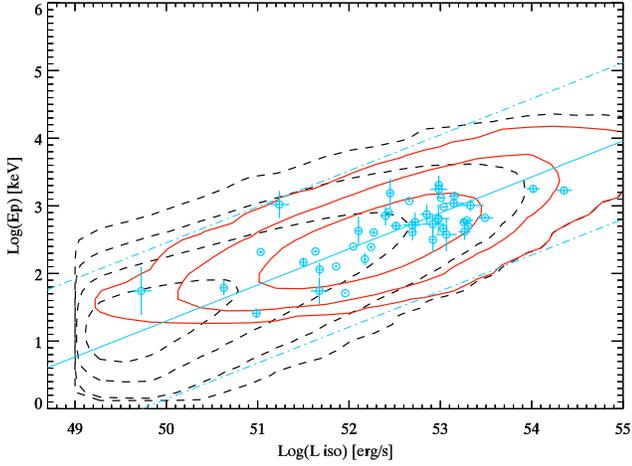,width=9.5cm}
\caption{
Simulation of GRBs with luminosity evolution (see Tab. \ref{tab1} for the parameters) 
assuming the real \yone\ correlation observed in the complete \sw\ sample of GRBs. 
Same symbols as in Fig. \ref{fg1}. }
\label{fg2a}
\end{figure}

\subsection{Real correlation}

 We repeat the Monte Carlo simulation to obtain a population of GRBs assuming that \ep\ and \liso\ are intrinsically correlated. We assume that the correlation between \ep\ and \liso\ has the same properties as the correlation that we observe in the complete sample of Swift bursts (point (3) of \S 2). In particular we refer to the slope, normalization and scatter found in N12.
Then, from our simulated population of GRBs, we select those that satisfy the requirements on the peak flux and on \epo\ and we study the \yone\ correlation in this sub--sample. 
In this sub--sample  
we expect, of course, to find a strong correlation, since we have assumed it in simulating the population of GRBs.

The main reason for performing this test is for self--consistency and also to verify wether the cuts introduced 
(flux limit and constraints on \epo) in defining the complete sample of simulated GRBs 
 do affect the characteristics of the correlation and lead us to find a correlation which is different from the correlation that we have assumed for the simulation. For example, if we simulate a sample satisfying a given correlation, we can expect that the introduction of a flux limit could reduce the scatter (and maybe also modify the slope) of the correlation that we find.
 

The results are shown in Fig. \ref{fg2a} and reported in Tab. \ref{tab1}. We find a good agreement between the complete \sw\ sample 
of simulated bursts (solid contours in Fig. \ref{fg2a}) and the complete \sw\ sample 
of real bursts (open points in Fig. \ref{fg2a}). 
All the repeated simulations (we have assumed the LF with luminosity evolution but 
the results do not depend on this choice) produce a significant correlation in 
the \yone\ plane (see Tab. \ref{tab1}) and in 66\% of the simulations such a significant 
correlation is consistent in slope, normalization and scatter with that defined by the \sw\ complete sample.

\subsection{Boundary}

\begin{figure}
\hskip -0.6truecm
\psfig{file=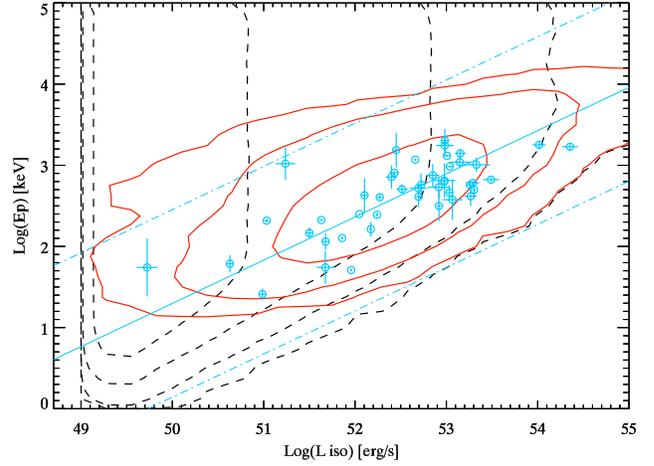,width=9.5cm}
\caption{
Simulation of GRBs with luminosity evolution (see Tab. \ref{tab1} for the parameters) assuming 
the real \yone\ correlation as a boundary: the distribution probability of \ep\ is from Eq. \ref{boundary}. 
Same symbols as in Fig. \ref{fg1}.  
}
\label{fg4}
\end{figure}

We can also test the possibility that the \yone\ correlation is  caused by a boundary. 
To this aim, we perform the simulation by assuming that bursts follow the boundary 
represented by the  $\log(E_{\rm peak})=m\log(L_{\rm iso})+q$ (with $q$ and $m$ being the normalization and 
slope of the \yone\ correlation found by N12 for the complete \sw\ sample). 
For each simulated value of \liso, we assign a peak energy \ep\ according to a probability distribution:
\begin{eqnarray}
\phi(E_{\rm peak})&=& A e^{-\frac{(E_{\rm peak}-E_{\rm peak,c})^2}{2\sigma^2}}; 
                      \,\,\,\, E_{\rm peak}\le E_{\rm peak,c}\nonumber \\
	              &=& A; \,\,\,\,\, \qquad \qquad \qquad \qquad E_{\rm peak}> E_{\rm peak,c} 
\label{boundary}
\end{eqnarray}
where  $\log(E_{\rm peak,c})=m\log(L_{\rm iso})+q$. 
This distribution introduces a sharp decrease of the simulated bursts below the observed \yone\ correlation, 
 which then forms a boundary in the \yone\ plane with a uniform distribution of GRBs above it. 
We report the results in Tab. \ref{tab1} and show them in Fig. \ref{fg4}. 
Under this hypothesis, we find that a considerably large  fraction (87\%) of the repeated 
simulations produces a strong \yone\ correlation but only  12\% of the correlations seem consistent with 
the correlation of the \sw\ complete sample. 1.4\%  of outliers at more than 3$\sigma$ above the correlation are found. 
By contrast, in the samples of bursts without measured redshifts of \ba\ and \fe, 
Nava et al. (2011) have shown that there are no outliers at more than 3$\sigma$ of the \yone\ correlation.

\section{Conclusions}

We performed Monte Carlo simulations of GRBs with a given redshift 
distribution and luminosity function (as described in Eq. \ref{z} and Eq. \ref{lf}, respectively). 
We have made the null hypothesis that there is no 
intrinsic correlation between the luminosity \liso\ and the peak energy \ep\ of the simulated bursts. 
We have considered the GRBs that have a peak flux $\ge$ \flim = 2.6 ph cm$^{-2}$ s$^{-1}$ which is 
the same flux limit of the complete \sw\ sample studied in S12 and N12 and we also required 
that \epo\ of the detected bursts can be measured by current instruments, i.e. that it 
lies in the 15 keV--2 MeV energy range. 
These are the simulated bursts that would be {\it detected} by \sw. 
The use of a flux--limited sample of \sw\ bursts as that defined in S12 has the advantage that it 
avoids several complexities related to the \sw\ trigger method (e.g. dependence from the time evolution of the signal and significance with respect to a time--variable background or dependence from off--axis detector response). In particular the flux--limit of the S12 sample is sufficiently high that 
 the selected sample should be  free from these trigger--related issues. 

If we make the hypothesis that there is no correlation between \ep\ and \liso, only in 7.3\% of the 
repeated simulations (e.g. for the case of a GRB population evolving in density with redshift, as found in S12) 
we find a statistically robust (i.e. chance probability of the rank correlation coefficient $\le 10^{-3}$)  \yone\ correlation. If we also 
require that our simulations  produce a correlation similar (in slope, normalization and scatter) to that observed among real GRBs of the complete \sw\ sample, the 
percentage reduces to 0.7\%. 
Therefore, we reject the null hypothesis that there is not an intrinsic \yone\ correlation 
at the 2.7$\sigma$ level of confidence (3.0$\sigma$ for the case of luminosity evolution). 
These results suggest that a correlation between \liso\ and \ep\ should exist. 
Since for \sw\ we have considered a bright cut on the peak flux (to match that adopted in 
the definition of the complete sample of S12), our results
are obtained for the most conservative case. 
Similar results, i.e. the impossibility to produce a strong \yone\ correlation as due to  
the trigger threshold, are also obtained for \ba.

An alternative possibility is that there is a boundary in the \yone\ plane (Nakar \& Piran 2005). 
If we assume that this boundary is coincident with the observed \yone\ correlation,
the scatter of the data points around it is not symmetric but there is a 
larger fraction of bursts with intermediate \ep\ and low luminosities. 
The trigger threshold here could play a role in hiding the events with extremely low luminosity. 
Our simulations, assuming the existence of the \yone\ correlation as a boundary, suggest that  
while a considerable percentage (87\%) of simulations produce a significant correlation 
in the \yone\ plane, only 12\% of the simulations also produce a correlation with slope, normalization and scatter 
consistent with that of the real GRB sample. This result excludes at the 1.6$\sigma$ confidence level the boundary case, as we have modeled it here. In particular, we 
assumed that the distribution of bursts above the boundary is uniform. It could still be possible, however, that a boundary with an asymmetric scatter below and above 
it exists due to some other property of GRBs (e.g. their jet opening angle). This  possibility requires more elaborated simulations that are outside the scope of this paper. 
In conclusion, our simulations show that there is a correlation between \ep\ and \liso\ and this cannot be due to a selection bias caused by a flux--limited sample.

\section*{Acknowledgments}
We acknowledge ASI (I/088/06/0  and I/004/11/0), a 2010 PRIN--INAF grant and PRIN--MIUR 2009 ERC3HT for
financial support. D. Burlon is thanked for comments.  We acknowledge the referee for useful comments.

\end{document}